\documentstyle[12pt]{article}

\begin{document}

\title{Relaxation dynamics in the presence of unequally spaced attractors along the
reaction coordinate}
\date{December 1, 2000}
\author{F. Despa and R.S. Berry \\
Department of Chemistry,\\
The University of Chicago,\\
Chicago, Illinois 60637}
\maketitle

\begin{abstract}
We show how reaction coordinate path lengths affect the relaxation
efficiency of a complex system. To this purpose, we consider the metric
contributions to the transition rates. These metric contributions preserve
informations about the geometry change of the system at the barrier crossing
and, therefore, are directly related to the path lengths. The output of the
present work can lead to identify a way to enrich our knowledge about the
ability of the complex systems to relax only to preferential structures.
\end{abstract}

The different aspects of the topographies and topologies of multidimensional
potential surfaces of polyatomic molecules, clusters, and nanoscale
particles and the way they govern the dynamics and phaselike behavior of
these systems are currently receiving a great interest; for reviews, see 
\cite{berry,ball,wales1} and the references therein. As a result, a new
picture of their energy landscape is emerging which greatly enhances our
understanding in this pressing issue. Despite the enormous power to do
computations and many methods for extracting useful information, the problem
of relating potential surfaces of complex systems to their relaxatio is far
from a complete solution.

For complex systems, development of better potential functions still remains
clearly a priority. We can say that, at least, starting from available
expressions for the interaction between and among the system constituents,
accurate potential energy surfaces (PES) may be constructed for many
systems. Finding minima and saddles on potential surfaces has become a
straightforward matter by employing some of the most widely used algorithms
or combinations thereof.\cite{press,wales,davis,hinde} As it evidenced the
problem shifted from how to find minima efficiently to what method should be
employed to provide a better connection between the topography and the
dynamics on the surface. Statistical methods,\cite{kunz,doye1,white}
especially that based on the topography diagnosis procedure,\cite{kunz}
achieved considerable success, caying the study of complex systems to the
point where one can begin to infer dynamics of flow on the surface from the
general characteristics of the topography. A serious shortcoming of the
existing dynamical studies, however, has to be pointed out: virtually, all
of them describe the thermally activated barrier crossing by the traditional
transition state theory (TST) rate model which deals only with the height of
the barrier and the densities of initial and saddle states.\cite{forst} No
information about the geometrical structures of the initial and saddle sites
in this approach. From this point of view, there is a limit to the precision
with which we can characterize the capability of complex systems to relax
preferentially to one of only a limited number of geometrical structures
from the vastly larger variety that the system might exhibit. However, there
are other ways to compensate for this lack of information. It was shown, for
example, that the topographical properties of the potential energy surface
(PES) determine the extent to which the system is either glass-forming or
focusing on regular structures.\cite{ball,kunz,miller} In addition,
qualitative interpretations of the interplay between structure and dynamics
have been made of the path length distributions between connected minima
along the reaction coordinate.\cite{miller,wales2}

For a better understanding of dynamics, it is now useful to investigate the
relation between the variety of path lengths between neighboring minima
along the reaction coordinate of a specific PES and the flow of probability
between those minima. We attempt in the following to get more insight into
this particular aspect by invoking a generalization of transition-state
theory (TST) to express the transition rates on the PES.\cite{hanggi} To
this purpose, we include the metric contributions to the transition rates
and speculate on their relation to the reaction coordinate paths brought
about in this way. We work out a simple three-level system with a model
surface based on a ''nearest-neighbor'' connection pattern and focus on the
ability of the system to find the global minimum under the assumption of
unequal path lengths between connected minima.\medskip 

Let $x$ denote the reaction coordinate of a complex system with $N$ degrees
of freedom whose dynamics is governed by an appropriate potential surface.
The reaction coordinate $x$ is a function of all the degrees of freedom $%
x=\left( q_{1},...q_{N};p_{1},...p_{N}\right) $, where $q_{i}$ stands for
the $i^{th}$ configuration coordinate and $p_{i}$ for its momentum. In the
following, we restrict our considerations to a small region of the PES
containing two states of local stability $\Theta _{1}$ and $\Theta _{2},$
for which the reaction coordinate takes the particular values $x_{1}$ and $%
x_{2}$ with the corresponding energy levels $E_{1}$ and $E_{2}$,
respectively. In general, these two domains of attraction might be separated
by barriers containing one or possibly more saddle points, possible unstable
limit cycles, or even more complex unstable attractors, including
combinations thereof. For most of the following we shall restrict our
discussion of the theory to the situation in which two adjacent attracting
basins are separated by a single saddle point. In the present case, we
assume that between the local minima $\Theta _{1}$ and $\Theta _{2}$ there
is a transition state at $x_{\left( 1-2\right) }\equiv x_{a}$.

Generally, the forward TST rate $w_{1-2}$ is given by\cite{chandler} 
\begin{equation}
w_{1-2}=\frac{\left\langle \delta \left( x-x_{a}\right) \stackrel{.}{x}%
\left( a\right) \theta \left[ \stackrel{.}{x}\left( a\right) \right]
\right\rangle }{\left\langle \theta \left( x_{a}-x\right) \right\rangle }%
\;\;,  \label{1}
\end{equation}
where $\delta $ and $\theta $ are the usual {\it delta}- and {\it step}%
-functions. Here, the average $\left\langle ...\right\rangle $ denotes an
equilibrium average over the canonical probability density. The integration
over the momenta is straightforward and leads to 
\begin{equation}
w_{1-2}=\left( 2\pi \beta \right) ^{-1/2}\frac{\left[ \delta \left(
x-x_{a}\right) \left| {\bf \nabla }_{Q}\left( x\right) \right| \right] }{%
\left[ \theta \left( x_{a}-x\right) \right] }\;\;,  \label{2}
\end{equation}
where $\left| {\bf \nabla }_{Q}\left( x\right) \right| ^{2}=\sum_{i}\left( 
\frac{\partial x}{\partial Q_{i}}\right) ^{2}$ and $Q_{i}$ are the
mass-weighted coordinates $\left( Q_{i}=q_{i}m_{i}^{1/2}\right) $. $\left[
...\right] $ indicates an average over the coordinates only. Further
simplifications in $\left( 2\right) $ can be achieved by integrating over
the coordinates of the center-of-mass position $R$ and all orientations $%
\Omega $ relative to a reference configuration of the system. This follows
the coordinate transforming $\left( Q_{1},...,Q_{N}\right) \rightarrow
\left( R,\Omega ;f_{1}...f_{m}\right) $, where ${\bf f}$ are the internal
coordinates (in number, $m=N-6$). The internal coordinates allow us to
describe forces acting in the system in terms of the potential function $%
U\left( {\bf f}\right) $. For example, the number of particle in the well $%
\left[ \theta \left( x_{a}-x\right) \right] $ can be written after suitable
integrations as 
\begin{equation}
\left[ \theta \left( x_{a}-x\right) \right] =\int df_{1}...df_{m}J\left( 
{\bf f}\right) \theta \left( x\left( {\bf f}\right) \right) \exp \left( -%
\frac{U\left( {\bf f}\right) }{k_{B}T}\right) \;\;.  \label{3}
\end{equation}
$J\left( {\bf f}\right) $ is the integrated Jacobian of the coordinate
transformation and comprises the determinant of the metric tensor times the
volume of the system and factors resulting from integrations over the
orientational degrees of freedom.

In the low temperature limit ($\beta U\left( f\right) \gg 1$, with $\beta =%
\frac{1}{k_{B}T}$), eq. $\left( 2\right) $ can be evaluated by a Gaussian
steepest-descent approximation. The procedure needs to expand the effective
potential $U_{eff}\left( {\bf f}\right) =U\left( {\bf f}\right) -k_{B}T\ln
J\left( {\bf f}\right) $ to second order at the saddle point $\left(
a\right) $ and at the local minimum $\left( \Theta _{1}\right) .$ After some
simple algebra, the final formula for the TST forward rate can be given in
the form 
\begin{equation}
w_{1-2}\simeq \frac{\left| {\bf M}^{-1/2}\right| }{2\pi }\left( \frac{J_{a}}{%
J_{1}}\right) \frac{\prod_{i=1}^{m}\lambda _{i}^{\left( 1\right) }}{%
\prod_{i=1}^{m-1}\lambda _{i}^{\left( a\right) }}\exp \left( -\beta
E_{1-a}\right) \;\;.  \label{4}
\end{equation}
Note that in the case of multiple transitions states we have to count each
contribution separately and write the rate as a sum over pathways.\cite{kunz}
The $\simeq $ sign is used instead of the equality sign because of terms of
order $O\left( \frac{1}{\beta {\bf f}}\right) $ compared to unity have been
neglected in above. ${\bf M}$ is the mass matrix and the indices $\left(
a\right) $ and $\left( 1\right) $ indicate that the corresponding quantities
are evaluated at the saddle point and the local minimum $1,$ respectively. $%
E_{1-a}$ measures the barrier height for the forward transition. $\lambda
_{i}$ are the eigen values of the force constant matrix, that is the {\it %
covariant} second derivative of the energy. The Jacobians $J_{1,a}$ in eq. $%
\left( 4\right) $ contain information about the volume, bond lengths and
orientations of the system at the local minima $\Theta _{1}$ and the saddle
point, respectively. Equivalently said, the metric contributions $J_{1,a}$
are directly related to the values of the reaction coordinate at these
particular sites the system is visiting in the evolution through the
configuration space. Under these circumstances, the above contributions
modify the rate constant by a multiplicative factor. This factor is the
ratio between the two Jacobians corresponding to the atomic arrangements at
the saddle point and at the position of the initial well, respectively, all
taken in the same reference frame. Within the Gaussian steepest-descent
approximation for integrations in the phase space, this Jacobian ratio can
be taken constant. for any set of coordinates employed. The backward rate $%
w_{2-1}$ may be obtained straightforwardly by replacing $J_{2}$ and $E_{2-a}$
for $J_{1}$ and $E_{1-a}$, respectively.

In the following we shall apply the TST rate formula to a simple landscape
case in which the two states of local stability $\Theta _{1}$ and $\Theta
_{2}$ are connected to the global minimum, $\Theta _{3}$. The global minimum
has the energy $E_{3}$ $\left( E_{3}<E_{1,2}\right) $ and is placed at $x_{3}
$ along the reaction coordinate. A barrier placed at $x_{\left( 2-3\right)
}\equiv x_{b}$ separates the global minimum from the nearest-neighbor
attractor $\Theta _{2}$. All the considerations in above apply equally to
the forward $\left( w_{2-3}\right) $ and backward $\left( w_{3-2}\right) $
transitions between the attractors $\Theta _{2}$ and $\Theta _{3}$. These
can be constructed in a similar manner by using the parameters $J_{b}$, $%
J_{2,3}$, $\lambda _{i}^{\left( b,2,3\right) }$ and $E_{2-b,3-b}$.

The general procedure describing the dynamics on the PES relies in practice
on the explicit knowledge of a master equation governing the time dependence
for the single-event probability $P_{i}\left( x_{i},t\right) $ of the
reaction coordinate. For the PES described above, the system of master
equations has the form 
\begin{eqnarray}
\frac{\partial P_{1}}{\partial t} &=&w_{2-1}P_{2}\left( x_{2}\right)
-w_{1-2}P_{1}\left( x_{1}\right)   \nonumber \\
\frac{\partial P_{2}}{\partial t} &=&-\left( w_{2-1}+w_{2-3}\right)
P_{2}\left( x_{2}\right) +w_{1-2}P_{1}\left( x_{1}\right)
+w_{3-2}P_{3}\left( x_{3}\right)   \label{5} \\
\frac{\partial P_{3}}{\partial t} &=&w_{2-3}P_{2}\left( x_{2}\right)
-w_{3-2}P_{3}\left( x_{3}\right) \;\;\;.  \nonumber
\end{eqnarray}
The practical problem is to determine the extent to which the dynamics of
the system is affected by the metric contributions $J$ to the corresponding
TST rates $w_{i-j}$ $\left( i,j=\overline{1,3}\right) .$ To do so , we have
to solve the above system of kinetic equations under certain initial
conditions. For simplicity, let us assume that at the initial moment $\left(
t=0\right) $ the system is in the domain of attraction $\Theta _{1}$ with
the probability $P_{1}\left( t=0\right) =1$. Here, the evolution of the
reaction coordinate through the configuration space starts with the highest
probability from $x=x_{1}$.

For the present case, analytic solutions of eqs. $\left( 5\right) $ are
available: 
\begin{eqnarray}
P_{1}\left( t\right)  &=&\alpha \left[ 1+\frac{g_{1}S_{6}}{2S_{5}S_{2}}\phi
\left( \frac{2+\chi }{\chi }\frac{\varphi }{\phi }e^{-\omega _{1}t}-\frac{%
2+\psi }{\psi }e^{-\omega _{2}t}\right) \right] \;\;,  \nonumber \\
P_{2}\left( t\right)  &=&\alpha g_{1}\left[ 1+\frac{S_{6}}{2S_{2}}\left( 
\frac{2+\chi }{\chi }e^{-\omega _{1}t}-\frac{2+\psi }{\psi }e^{-\omega
_{2}t}\right) \right] \;\;,  \label{6} \\
P_{3}\left( t\right)  &=&\alpha g_{1}g_{3}\left[ 1+\frac{1}{2S_{2}}\left[
\left( 2+\chi \right) e^{-\omega _{1}t}-\left( 2+\psi \right) e^{-\omega
_{2}t}\right] \right] \;\;,  \nonumber
\end{eqnarray}
which were obtained in terms of eigenvalues and eigenvectors of the
characteristic system of equations. The (nontrivial) eigenvalues are given
by $\omega _{1,2}=\frac{w_{2-1}}{2}\left( S_{1}\pm S_{2}\right) $ with $%
S_{1}=1+g_{1}+g_{2}+g_{3}$, $S_{2}=\sqrt{\left( 1+g_{1}-g_{2}-g_{3}\right)
^{2}+4g_{3}}$, $g_{1}=\frac{w_{1-2}}{w_{2-1}}$, $g_{2}=\frac{w_{3-2}}{w_{2-1}%
}$ and $g_{3}=\frac{w_{2-3}}{w_{2-1}}$. All the other constants entering eq. 
$\left( 7\right) $ are given by $\chi =S_{4}-S_{2}$, $\phi =S_{2}+S_{3}$, $%
\varphi =S_{2}-S_{3}$, $\psi =S_{4}+S_{2}$ and $\alpha =\frac{S_{5}}{S_{6}}%
\left[ \left( g_{1}+\frac{S_{3}-S_{2}}{2}\right) \frac{S_{4}-S_{3}}{S_{6}}%
+\left( g_{1}-\frac{S_{3}-S_{2}}{S_{4}-S_{2}}\right) \right] ^{-1}$with $%
S_{3,4}=1\mp g_{1}\pm g_{2}+g_{3}$, $S_{5}=S_{3}^{2}-S_{2}^{2}$ and $%
S_{6}=S_{2}^{2}-S_{4}^{2}$. 

The energy levels corresponding to the minima $\Theta _{1}$, $\Theta _{2}$
and $\Theta _{3}$ of the standard potential energy surface employed in the
present study are disposed as in a steeper funnel obeying the following
sequence $E_{1}>E_{2}>E_{3}$. Each transition state connecting two adjacent
minima lies an energy $l_{i}$ above the nearest uphill minimum. Therefore,
one sets $E_{2-a}-E_{1-a}=l_{1}>0$ and $E_{3-b}-E_{2-a}=l_{2}>0$,
respectively. In addition, we assume in this model that the potential energy
barrier opposing the escape from the global minimum is higher than that
opposing the escape from the minimum $\Theta _{2}$ towards the minimum $%
\Theta _{1}$, $\left( E_{3-b}>E_{2-a}\right) $. It is also assumed that the
relaxation rate from $\Theta _{2}$ towards the global minimum is
energetically greater than the uphill escape $E_{2-a},$ $\left(
E_{2-a}-E_{2-b}=l_{3}>0\right) $. In our computation $\lambda ^{\left(
1-3\right) }$ denote the vibrational frequencies of the reaction coordinate
in the corresponding minima and we shall keep them at constant values. As
for the mean vibrational frequency of the transition state, we make the
usual assumption,\cite{doye} that is the geometric mean of the vibrational
frequencies of the two minima it connects, $\lambda ^{\left( a\right) }=%
\sqrt{\lambda ^{\left( 1\right) }\lambda ^{\left( 2\right) }}$ and $\lambda
^{\left( b\right) }=\sqrt{\lambda ^{\left( 2\right) }\lambda ^{\left(
3\right) }}$, respectively. 

By using probability distributions $P_{i}$ $\left( i=1,2,3\right) $ as
derived above, we have investigated the tendency of relaxation of the
complex system under various circumstances for reaction coordinate paths. In
simulating the distribution of path lengths we speculate on the direct
relation between the Jacobian $J\left( {\bf f}\right) $ and the reaction
coordinate, as can be seen below. The results are displayed in Fig.1.

Going back to the purpose of the present paper, we focus on the role the
metric contribution plays in the relaxation behavior of the system and
assign to the Jacobians $J_{1}$, $J_{2}$, $J_{3}$, $J_{a}$ and $J_{b}$ the
following sequence of numbers $J_{1}:J_{2}:J_{3}\equiv
1:2:3,\;J_{a}:J_{b}\equiv 4:5$. Roughly, these numbers should correspond to
a distribution of path lengths along the reaction coordinate with increasing
step sizes towards the global minimum. The pictorial correspondence in
nuclear configuration space is therefore characterized by sizeable
rearrangements of the system components on their way to relax towards the
ground state. We are interested in monitoring the relaxation dynamics of the
system on the particular energy landscape as described in above. The values
of the parameters $l_{1-3}$ have been tuned to achieve a rapid saturation
effect on a scale of $100$. The time evolution of the probability $P_{3}$
for which the system relaxes to the global minimum is displayed in Fig. 1
(see the curve labeled $\left( 1\right) $). One can observe that the
increase in the population of the ground state is sharp and the probability
reaches a plateau. We now modify the metric contributions to the relaxation
rates by considering $J_{1}\equiv J_{2}\equiv J_{3}\equiv AJ_{a}\equiv
BJ_{b}=1$ which corresponds to an equally spaced distribution of minima
along the reaction coordinate. All the other parameters in above remain at
constant values. Looking at Fig. 1 (see the curve $2$), we observe that the
accumulation in the global minimum is much slower in this case. The time
evolution of the probability $P_{3}$ does not reach saturation on the same
time scale. This indicates that the backward rates of escape from attraction
domains are higher for the present case in comparison with the former and,
the system spends considerably more time now moving uphill on the PES. The
geometrical structures corresponding to minima $\Theta _{1}$ and $\Theta _{2}
$ attract the system at rates comparable to that of the ground state
structure even if the latter is energetically more favorable. Note that, the
balance may decisively be turned around if, by chance, the numbers of
pathways towards $\Theta _{1}$ and $\Theta _{2}$, respectively, are large
enough to compensate the energy gaps by entropic contributions.\cite{kunz}
The efficiency of relaxation towards the global minimum can be reduced
dramatically by inverting the numbers in the sequence $J_{1}:J_{2}:J_{3}$
from those assumed in the first example. This becomes $J_{1}:J_{2}:J_{3}%
\equiv 3:2:1$, and the pictorial correspondence of the PES may be that of a
steeper funnel with a wide step at the top, between the minima $\Theta _{1}$
and $\Theta _{2}$ and a narrow one at the bottom, between the minima $\Theta
_{2}$ and $\Theta _{3}$. (We assume $J_{a}:J_{b}\equiv 1,\;$for the present
case). As can be seen in Fig.1 (see the curve $3$), the probability $P_{3}$
of accumulation in the global minimum is even worse than that of the
previous case. In turn, a narrow barrier between the global minimum $\Theta
_{3}$ and adjacent local minimum $\Theta _{2}$ combined with a broader one
between $\Theta _{2}$ and $\Theta _{3}$ should result in a longer survival
of the system in the well around energy level $E_{2}$. This is demonstrated
by curve $3^{\prime }$ in Fig.1 which shows that the accumulation in the
domain of attraction $\Theta _{2}$ is still rising over the entire time
scale for the above values of the metric contributions. This behavior
contrasts with the two previous situations (see curves $1^{\prime }$ and $%
2^{\prime }$ in Fig.1) where, after an initial increase in population, the
attractor $\Theta _{2}$ starts, more or less suddenly, to depopulate.

In conclusion, we can say that the efficiency of relaxation towards the
global minimum can be much affected by the metric contributions to the
transition rates. The metric contributions are directly related to the
distribution of the path lengths by the integrated Jacobians. Therefore,
these play, in concert with the barrier heights, an implicit role in
classifying the archetypal energy landscapes.\cite{ball,wales1,kunz,miller}
In addition, we can say that incorporating the metric contribution to the
TST rate is the appropriate way to combine the height of the barrier and the
densities of initial and saddle states with the differential path length
corresponding to the transition of the system between these specific
stationary points. The present approach allows the TST rate to be implicitly
related to the geometrical structures involved in the min-saddle-min
transition of the system. We intend to use the information so obtained to
identify a way to enrich the precision with which we can characterize the
ability of complex systems to relax, preferentially, to only a limited
number of geometrical structures from the vastly larger variety that the
system might exhibit.

{\Large Figure captions}

\medskip

Fig. 1 - Time evolution of the probability $P_{3}$ (curves $1$, $2$ and $3$)
and $P_{2}$ (curves $1^{\prime }$, $2^{\prime }$ and $3^{\prime }$) for
three different sequences of path lengths between connected minima. The
initial population in the highest minimum was assumed equal to unity, $%
P_{1}\left( t=0\right) =1$. For explanation, see the text.

\end{document}